\documentstyle[12pt]{article}
\input epsf
\begin{document}
\bibliographystyle{unsrt}
\def\question#1{{{\marginpar{\small \sc #1}}}}
\newcommand{\bra}[1]{\left < \halfthin #1 \right |\halfthin}
\newcommand{\ket}[1]{\left | \halfthin #1 \halfthin \right >}
\newcommand{\be}{\begin{equation}}
\newcommand{\ee}{\end{equation}}
\newcommand{\vsig}{\mbox {\boldmath $\sigma$\unboldmath}}
\newcommand{\vep}{\mbox {\boldmath $\epsilon$\unboldmath}}
\newcommand{\fn}{\frac 1{E^i+M_N}}
\newcommand{\fs}{\frac 1{E^f+M_N}}
\newcommand{\gamQQ}{\Gamma({Q \bar{Q}}_V \rightarrow \gamma + R)}
\newcommand{\brQQ}{b_{rad}({Q \bar{Q}}_V \rightarrow \gamma + R)}
\newcommand{\gamRgg}{\Gamma(R \rightarrow gg )}
\newcommand{\bRgg}{br(R \rightarrow gg )}
\newcommand{\qqbar}{$q \bar{q}~$}
\newcommand{\gsi}{\,\raisebox{-0.13cm}{$\stackrel{\textstyle>}
{\textstyle\sim}$}\,}
\newcommand{\lsi}{\,\raisebox{-0.13cm}{$\stackrel{\textstyle<}
{\textstyle\sim}$}\,} 

\rightline{RU-96-92}  
\baselineskip=18pt
\vskip 1.7in
\begin{center}
{\bf \LARGE Determining the Gluonic Content of Isoscalar Mesons}\\
\vspace*{0.9in}
{\large Glennys R. Farrar}\footnote{Invited talk ICHEP96, Warsaw, Poland,
July 29, 1996 based on work done in collaboration with M. B. Cakir, F. E.
Close and Z. P. Li.} \\ 
{\it Department of Physics and Astronomy}\\
{\it Rutgers University, Piscataway, NJ 08855, USA}\\
\vspace*{0.7in} 
\end{center}


\begin{abstract} 
The gluonic widths of four leading glueball candidates are determined 
from their production in radiative quarkonium decays, allowing
quantitative estimation of their glue content.  Lattice predictions
for the scalar and tensor channels seem to be in reasonable agareement
with present data (allowing for mixing with $q \bar{q}$ states).
However there is a glueball-like-state in the pseudoscalar spectrum
whose mass is considerably lower than expected from lattice estimates.
\end{abstract}

\newpage

This talk summarizes some recent results of Close, Farrar, and
Li\cite{f:xx} (CFL), in which quantitative methods for determining the
gluonic content of isosinglet mesons are utilized to study four states
which are promising glueball candidates\footnote{In additional work
not reported here, ref. \cite{f:xx} analyzes two generic three-state
mixing scenarios for the $f_0$ sector, gives the constraints which
follow from knowledge of $\gamma \gamma$ widths, and tests validity of
the procedures on known $q \bar{q}$ resonances.  The reader is refered
to ref. \cite{f:xx} for most references and details; only the
essentials are given here.}:  
\begin{itemize}
\item 
$f_0(1500)$
\item
$f_J(1710)$ where $J=0$ or $2$
\item $\xi(2230)$\cite{beijing} 
\item $\eta(1440)$, now resolved into two pseudoscalars.
\end{itemize}
These states satisfy qualitative criteria expected for
glueballs\cite{closerev}: 
\begin{enumerate}
\item
Glueballs should be favoured over ordinary mesons in the central
region of high energy scattering processes, away from beam and 
target quarks.  The $f_J(1710)$ and possibly the $f_0(1500)$ have been
seen in the central region in $pp$ collisions.
\item
Glueballs should be produced in proton-antiproton annihilation, where
the destruction of quarks creates opportunity for gluons to be
manifested.  This is the Crystal Barrel and E760 production mechanism, 
in which detailed decay systematics of $f_0(1500)$ have been studied.
The empirical situation with regard to $f_J(1710)$ and $\xi(2230)$ is
currently under investigation. The $\eta(1440)$ is clearly seen in
$p\bar{p}$ annihilation
\item
Glueballs should be enhanced compared to ordinary mesons in radiative
quarkonium decay.  In fact, all four of these resonances are produced
in radiative $J/\psi$ decay at a level typically of $\sim1$ part per
thousand. A major purpose of this paper is to decide whether these rates
indicate that these resonances are glueballs, or not. 
\end{enumerate}
Furthermore, lattice QCD predicts that the lightest ``ideal" (i.e.,
quenched approximation) glueball be $0^{++}$, with state-of-the-art
mass predictions of $1.55 \pm 0.05$ GeV\cite{ukqcd} and $1.74 \pm
0.07$ GeV\cite{weing}.  The tensor glueball is predicted to lie in the
2.2 GeV region.  However both lattice and sum rule calculations place
the lightest $0^{-+}$ glueball at or above the $2^{++}$ glueball so
that the appearance of a glueball-like pseudoscalar in the 1.4-1.5 GeV
region is unexpected.  See \cite{ct96} for a review of lattice
glueball predictions.

CFL reformulate and apply the relationship proposed by Cakir and
Farrar\cite{cak} (CF) between the branching fraction for a resonance
$R$ in radiative quarkonium decay, $b_{rad}({Q \bar{Q}}_V \rightarrow
\gamma +R) \equiv \Gamma({Q \bar{Q}}_V \rightarrow \gamma + X)$ and
its branching fraction to gluons, $br(R \rightarrow gg) \equiv \Gamma
(R \rightarrow gg ) / \Gamma(R\rightarrow \rm{all})$:  
\begin{eqnarray}
\label{CF}
b_{rad}(Q\bar Q_V\to \gamma +R_J) & = & \nonumber \\
\frac {c_Rx|H_{J}(x)|^2}{8\pi(\pi^2-9)}\frac{m_R}{M_V^2}\Gamma_{tot} br(R_J
\rightarrow gg),
\end{eqnarray}
where $M_V$ and $m_R$ are masses of the initial and final resonances, and
$x \equiv 1-\frac {m_R^2}{M^2_V}$; $c_R$ is a numerical factor and
$H_J(x)$ a loop integral.  For a resonance of known mass, total width
($\Gamma_{tot}$), and $J^{PC}$, a relationship such as eq. (\ref{CF}) would
determine $br(R \rightarrow gg)$ if $b_{rad}({Q \bar{Q}}_V \rightarrow \gamma
+R)$ were known.  CF argued that one expects $br(R[q \bar{q}]
\rightarrow gg) = 0(\alpha^2_s) \simeq 0.1-0.2$ and $br(R[G]
\rightarrow gg) \simeq O(1)$, providing quantitative information on
the glueball content of a particular resonance.  Using $H_J(x)$ determined in 
the non-relativistic quark model (NRQM), CF found that known $q\bar{q}$
resonances (such as $f_2$(1270)) satisfy the former and noted that the
$f_0(1710)$ might be an example of the latter.  CFL clarified the CF
relationship between $b_{rad}({Q \bar{Q}}_V \rightarrow \gamma +R)$
and $br(R \rightarrow gg)$, examining its dependence on the $<g g |R>$
form factors and the theoretical and experimental constraints on these
form factors.  CFL concluded that the CF relation can be used, possibly
with generalized $H_J(x)$ functions, for glueballs and light-$q
\bar{q}~$ mesons as well as for heavy $q \bar{q}~$ mesons.   

In the $x$ regime of immediate interest, $x \sim 0.5 - 0.75$,
$\frac{x|H_J|^2}{30-45} \sim O(1)$.  This enables eq. (\ref{CF}) to be
manipulated into a scaled form that exhibits the phenomenological 
implications immediately.  Specifically\cite{f:xx}:  
\begin{eqnarray}
10^3  br(J/\psi \rightarrow \gamma [0^{++},~2^{++},~0^{-+}]) & = &
\nonumber \\
(\frac{m}{1.5\; {\rm GeV}}) (\frac{\Gamma_{R\rightarrow
gg}}{[96,~26,~50]~ {\rm MeV}})  \frac{x|H_J(x)|^2}{35}. 
\end{eqnarray}
Having scaled the expression this way, because $\frac{x|H_J|^2}{30-45}
\sim O(1)$ in the $x$ range relevant for production of 1.3 - 2.2 GeV
states, we see immediately that the magnitudes of the branching  ratios are
driven by the denominators 96 and 26 MeV for $0^{++}$ and $2^{++}$, and $50$
MeV for $0^{-+}$.   Thus if a state $R_J$ is produced in $J/\psi \rightarrow
\gamma X$ at $O(10^{-3})$ then $\Gamma (R_J \rightarrow gg)$ will typically
be of the order $100$ MeV for $ 0^{++}$, $O(25 ~ {\rm MeV})$ for
$2^{++}$, and $O(50 ~ {\rm MeV})$ for $0^{-+}$. 

We now apply eq. (2) to the glueball candidates listed above to infer
$\Gamma(R \to gg)$ for each of them.  Dividing by their measured total
width then yields $br(R \to gg)$.  Let us begin with the established
scalar meson, $f_0(1500)$.  Using the branching fractions of Bugg et
al\cite{bugg}, gives\cite{f:xx} $br(f_0(1500) \rightarrow gg) \ge
0.9 \pm 0.2$.  This is significantly larger than the $O(\alpha_s^2)$
which would be expected for a pure $q\bar{q}$ system, and supports
this state as a glueball candidate.  On the other hand using BES
results reported at this conference\cite{landua}, implies\cite{f:xx}
$br(f_0(1500) \rightarrow gg) = 0.3-0.5$.  The interpretation of this
state cannot be settled until the experimental situation clarifies. 

The conclusions regarding the gluonic content of $f_J(1710)$ depend
critically on whether $J=0$ or $J=2$.  In the latter case, this analysis
indicates it is a clear $q \bar{q}$ system\cite{f:xx}.  However if 
$f_J(1710)$ is a single state with $J=0$, 
$br(f_0(1710) \rightarrow gg) \ge 0.52 \pm 0.07$,
in accord with fig. 14 of ref.\cite{cak}.  In this case the $f_0(1710)$ would
be a strong candidate for a scalar glueball.  Knowing the spin and $K \bar{K}$
branching fraction of $f_J(1710)$ is of great importance for a more detailed
quantitative understanding of the composition of this state.  

As noted above, lattice QCD predicts the ground state glueball is a
$0^{++}$ with mass $\approx 1.65 \pm 0.1$ GeV.  An interesting
possibility is that three $f_0$'s in the $1.4-1.7$ GeV region are
admixtures of the three isosinglet states $gg$, $s\bar s$, and $n\bar
n$\cite{cafe95}.  Recently there have been two specific schemes
proposed which are based on lattice QCD\cite{wein96} and the emergent
phenomenology of scalar  mesons\cite{cafe95}.  Based on the
preliminary data available these mixing schemes seem satisfactory, and
the lattice prediction is consistent with the observed spectrum.  In
\cite{f:xx}, a simplified formalism for treating a three component
system of this type is presented, making use of $\gamma \gamma$ widths
to constrain the system.  Further data on $\gamma \gamma$ is needed.

Now we turn to the narrow state $\xi(2230)$ observed in $J/\psi
\rightarrow \gamma \pi^+ \pi^-; \gamma K^+ K^-;$ $ \gamma K_s^o
K_s^o;$ $\gamma p \bar{p}$\cite{beijing}.  If evidence for this state
survives increases in statistics, the case $J=2$ would be consistent
with $\xi(2230)$ being a tensor glueball, while $J=0$ is inconsistent
with unitarity.  This would have significant implications for the
emergence of a glueball spectroscopy in accord with lattice QCD, which
predicted the tensor state at about 2.2 GeV\cite{ct96}. It would also
raise tantalising questions about the $0^{-+}$ sector, as follows. 

Many years ago, production of the ``$\eta(1440)$'' was observed to be
prominent in radiative $J/\psi$ decay, causing it to be identified as
a potential pseudoscalar glueball.  Subsequently it was realised that
multiple isosinglet states are contained in the 1.4-1.8 GeV region.
However the analyses of radiative $J/\psi$ decay indicating the
existence of additional structure were not in agreement. Using recent
data in $p\bar{p} \to \eta(1440) + \cdots $, ref. \cite{f:xx}
identifies the problematic measurement and proposes a consistent
picture that experiments should now pursue.  The tentative
interpretation of the pseudoscalar states is\cite{f:xx}:  
\begin{eqnarray}
\label{pseudo}
\eta(1295) \sim \eta^{n\bar{n}}; ~br_{gg} \sim O(\alpha_s^2) \sim 0.25 
\nonumber \\
\eta_L(1410) \sim G(+{q\bar{q}}); ~br_{gg} \sim 1 \nonumber \\
\eta_H(1480) \sim \eta^{s\bar{s}}(+G); ~br_{gg} \sim 0.5 
\end{eqnarray}
These conclusions can be sharpened if the widths and decays from
Crystal Barrel and Obelix converge, and if $J/\psi \to \gamma 0^{-+}$
is pursued further\cite{f:xx}.  

The experimental data on $0^{-+}$ production in radiative $J/\psi$
decays in this mass region need clarification before strong
conclusions can be drawn, but if the existence of two states in
the $1400 -1500$ MeV range, and their relative production 
(one or both much more strongly produced than $\eta(1295)$) is
confirmed, we have a serious challenge to theoretical expectations.
The experiments would appear to be telling us that the lightest
pseudoscalar glueball is much lighter than predicted in quenched
lattice QCD, ($2.16 \pm 0.27$ GeV\cite{ct96}).  In view of the naive
argument that unquenching will increase the mass of the $0^{-+}$
glueball due to level repulsion from the lower-lying $q \bar{q}$ state
$\eta'$ and the apparent possible success (within uncertainties noted
above) of the lattice QCD predictions for the $0^{++}$ and $2^{++}$
glueball masses, such a discrepancy between lattice QCD and nature
would be of great interest.  

A very speculative explanation for the mass and properties of the
$\eta(1410)$ is that it the gluino-gluino bound state\cite{gluino,cak}
predicted in certain SUSY-breaking models, possibly mixed with nearby
pseudoscalar $q \bar{q}$ states.  If nature were supersymmetric and
SUSY breaking did not violate $R$-invariance, the gluino mass would be
$O(100)$ MeV\cite{f:99,f:110}.  In that case the $0^{++}$ glueball
would be approximately degenerate with the pseudoscalar gluino-gluino
($\tilde{g} \tilde{g}$) and spin-1/2 gluon-gluino bound states.  This
would lead to an ``extra'' isosinglet pseudoscalar in the spectrum,
with mass around 1 1/2 GeV.  The pseudoscalar glueball expected in
conventional QCD would also be present, but at $\sim 2.2$ GeV
according to the lattice prediction.  Decay of a $\tilde{g} \tilde{g}$
system would necessarily go through gluons, since (i) its direct
couplings to quarks would be suppressed by heavy squark masses and
(ii) hadrons containing a single gluino would be too massive to be
pair produced by the decaying $\tilde{g} \tilde{g}$.  Thus
$br(\tilde{g}\tilde{g} \to gg) \sim 1$ as appears to be the case for
the $\eta(1410)$.   

In summary, the analysis of Close, Farrar, and Li\cite{f:xx} shows: 
\begin{itemize}
\item  The $f_0(1500)$ is at least half-glueball if the Bugg et al
analysis\cite{bugg} of the $4 \pi$ channel is confirmed,  but is less
so according to the BES results.  Analysis of MarkIII data on $J/\psi \to
\gamma \pi \pi$ is urgently needed.  At this moment the experimental
determinations of $\Gamma(J/\psi \to \gamma f_0(1500))$ are inconsistent.
\item  The $f_J(1710)$ is also at least half-glueball, if $J=0$; if 
 $J=2$ it is a $q \bar{q}~$ meson.  Experimental determinations of the
$f_{0,2}$ spectra in the $1.6-1.8$ GeV region are presently inconsistent.
\item  The $\xi(2330)$ is unlikely to have $J=0$, if present
experimental data are correct.  If it has $J=2$ it strongly resembles
a glueball.
\item The $\eta(1440)$ is separated into two states. The lower
mass state, $\eta(1410)$, has strong affinity for glue; the higher mass
$\eta(1480)$ is consistent with being the $s\bar{s}$ member of a
nonet, perhaps mixed with glue.
\end{itemize}
It is of urgent importance to (a) arrive at an experimental consensus
on the $f_0$ and $f_2$ masses and widths in the 1600-1800 region and
(b) resolve the discrepancies in the present determinations of
$br(\psi \to \gamma f_0(1500)) $.  Measurement of production branching
fractions of the $f_0$ and $f_2$ mesons in $\Upsilon$ radiative decay
should be quite easy and yield useful additional information.
Ref. \cite{f:xx} outlines a procedure to use data on $\psi \to \gamma
R$ and $\gamma \gamma \to R$ together, to help unravel the $q \bar{q}$ and
$gg$ composition of mesons.  To accomplish this, measurement of
$\Gamma(f_0(1370; 1500; 1710) \to \gamma \gamma)$ is an essential
ingredient. 

Experimental study of the isosinglet mesons does not yet allow a
definitive description of the spectrum in terms of the underlying QCD
quark and gluon degrees of freedom, and there are experimental
inconsistencies which need to be resolved.  However the emerging
picture of the scalar and tensor mesons is in remarkably good agreement
with quenched lattice QCD calculations.  On the other hand the
situation in the pseudoscalar channel is troublesome:  a glueball-like
state is clearly observed at about 1.4 GeV, which is far below the
quenched lattice QCD prediction.  Thus the next few years promise to
be very interesting for meson spectroscopy and QCD theory.
\section*{Acknowledgments}
It is a pleasure to thank Frank Close and Z.P. Li for our stimulating
and productive collaboration.

\end{document}